# Evolution of Gi-Fi Technology over other Technologies 

[1]Jyoti Tewari, [2]Swati Arya

[1] College of Engineering, Teerthankar Mahaveer University
Moradabad, Uttarpradesh, India

[2] College of Enginnering, Teerthankar Mahaveer University
Moradabad, Uttarpradesh, India

### Abstract

Gi-Fi stands for Gigabit Wireless. Gi-Fi is a wireless transmission system which is ten times faster than other technology and its chip delivers short-range multigigabit data transfer in a local environment. Gi-Fi is a wireless technology which promises high speed short range data transfers with speeds of up to 5 Gbps within a range of 10 meters. The Gi-Fi operates on the 60GHz frequency band. This frequency band is currently mostly unused. It is manufactured using (CMOS) technology. This wireless technology named as Gi-Fi. The benefits and features of this new technology can be helpful for use in development of the next generation of devices and places. In this paper, the comparison is perform between Gi-Fi and some of existing technologies with very high speed large files transfers within seconds it is expected that Gi-Fi to be the preferred wireless technology used in home and office of future.

## 1. Introduction

Gigabit Wireless is the world's first transceiver integrated on a single chip that operates at 60GHz on the CMOS (complementary metal–oxide–semiconductor) process. It will allow wireless transfer of audio and video data upto 5 gigabits per second, ten times the current Maximum wireless transfer rate, at one-tenth of the cost, usually within a range of 10 meters. In fact, GiFi is a wireless transmission system which is ten times faster than Wi-Fi and it is expected revolution networking in offices and homes by implementing high-speed wireless environments. It utilizes a 5mm square chip and a 1mm wide antenna burning less than 2milli watts of power to transmit data wirelessly over short distances, much like Bluetooth. Gi-Fi technology provides many features such as ease of deployment, small form factor, enabling the future of information management, high speed of data transfer, low power consumption etc. With growing consumer adoption of High-Definition (HD) television, low cost chip and other interesting features and benefits of this new technology it can be predicted that the anticipated worldwide market for this technology is vast. The new technology is predicted to revolutionize the way household gadgets talk to each other.

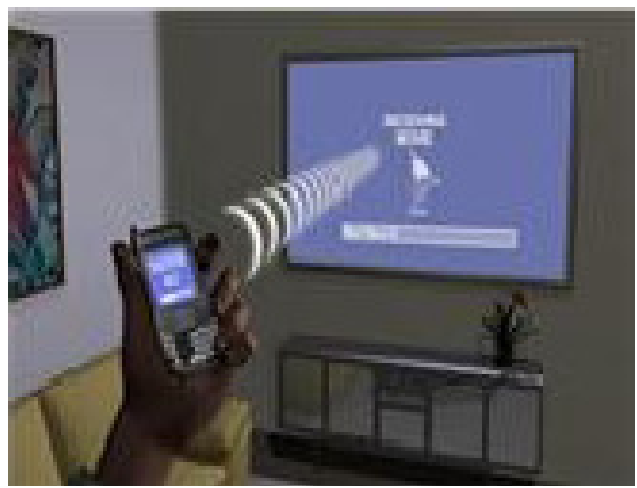

Fig.1 High speed local data transmission

Gi-Fi can be considered as a challenger to Bluetooth rather than Wi-Fi and could find applications ranging from new mobile phones to consumer electronics. Gi-Fi allows a full-length high definition movie to be transferred between two devices in seconds. to the higher megapixel count on our cameras, the increased bit rate on our music files, the higher resolution of our video files. Within five years, we expect Gi-Fi to be the dominant technology for wireless networking. By that time it will be fully mobile, as well as providing low-cost, high broadband access, with very high speed large files swapped within seconds which will develop wireless home and office of future. Gi-Fi potentially can bring wireless broadband to the enterprise in an entirely new way. Enhancements to next generation gaming technology is one of the other benefits of this technology.

The Nitro chipset in Gi-Fi technology by offering reduced size and power consumption, can be used to send and receive large amounts of data in a variety of applications, it is able to transfer gigabits of data within seconds and therefore it can be used for huge data file transmission and it is expected that this chipset replaces HDMI (High-Definition Multimedia Interface) cables and could develop wireless home and office of future.

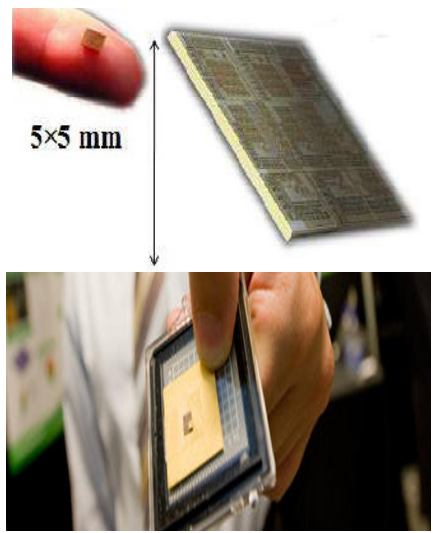

Fig.2 GI-FI chip

The GiFi chip is a good news for personal area networking because there is no internet infrastructure available to cop it with. It can have a span of 10 meters. The usable prototype may be less than a year away. With the help of gifi chips the videos sharing can be possible without any hurdles. The GiFi chip is one of Australia's most lucrative technologies.

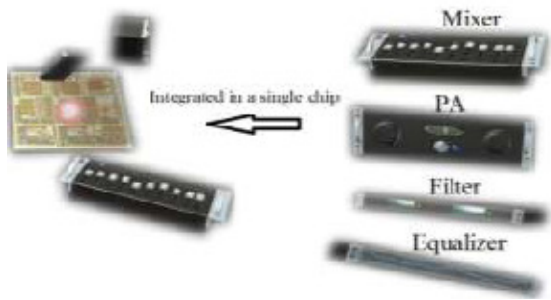

Fig.3 GI-FI chip in mobile devices

The new gigabit wireless system provides Multi-gigabit wireless technology that removes the need for cables between consumer electronic devices and is More than 100 times faster than current short-range wireless technologies such as Bluetooth and Wi-Fi. This technology with high level of frequency re-use can satisfy the communication needs of multiple customers within a small geographic region.

## 2. Gigabit Wireless Features

This Gi-Fi technology allows wireless uncompressed high-definition content and operates over a range of 10 meters without interference. Gi-fi chip has flexible architecture. It is highly portable and can be constructed in everywhere. Entire transmission system can be built on a cost effective single silicon chip that operates in the unlicensed, 57-64 GHz spectrum band. Gi-Fi technology also enables the future of information management, is easy to deployment with the small form factor.

### 2.1 Capacity of High Speed Data Transfer

The data transfer rate of Gigabit wireless technology is in Gigabits per second. Speed of Gi-Fi is 5 Gbps; which is 10 times the data transfer of the existing technologies. Providing higher data transfer rate is the main invention of Gi-Fi. An entire High-Definition (HD) movie could be transmitted to a mobile phone in a few seconds, and the phone could then upload the movie to a home computer or screen at the same speed.

### 2.2 Interference in Data Transfer

It uses the 60GHz millimeter wave spectrum to transmit the data, which gives it an advantage over Wi-Fi. Wi-Fi's part of the spectrum is increasingly crowded, sharing the waves with devices such as cordless phones, which leads to interference and slower speeds. But the millimeter wave spectrum (30 to 300 GHz) is almost unoccupied, and the new chip is potentially hundreds of times faster than the average home Wi-Fi technology.

### 2.3 Power Consumption

Power consumption of the present technologies such as Wi-Fi and Bluetooth are 5mili watts and 10mili watts but chip of Gi-Fi uses a tiny one-millimeter-wide antenna and it has less than 2mili watts of power consumption that in compare to the current technologies is very less.

### 2.4 Provides High Security

Gi-Fi technology is based on IEEE 802.15.3C and this standard provides more security since it provides optional security in the link level and service level. Point-to-point wireless systems operating at 60 GHz have been used for many years by the intelligence community for high security communications and by the military for satellite-to satellite communications.

## 3. Applications of GI–FI Technology

1) Gi-Fi technology has many attractive features that make it suitable for use in many places and devices. Gi-Fi technology offering reduced the chip size and power consumption, can be used to send and receive large amounts of data in a variety of applications For example, it is intended for use

   in a wide range of devices including personal computers, tablets, and smart phones. The technology's fast data-synchronization rates enable the rapid transfer of video, bringing the wireless office‖ closer to reality.

2) This technology can be effectively used in wireless pan networks, Inter-vehicle communication systems, Ad-hoc information distribution with Point-to-Point network extension, media access control (MAC), imaging and other applications.

3) Gi-Fi technology is able to transfer gigabits of data within seconds and therefore it can be used for huge data file transmission and it is expected that this chipset replaces HDMI cables and could develop wireless home and office of future.

4) Gi-Fi technology also can be used in broadcasting video signal transmission system in sports stadiums and mm-Wave video video-signals transmission systems. The technology could also be used for beaming full HD video in real-time and could be used by notebooks and other computers to wirelessly connect virtually all the expansion needed for a docking station, including a secondary display and storage.

## 4. Results

In recent years, new wireless local area networks (WLANs) such as Wi-Fi and wireless personal area networks (WPAN) such as Bluetooth have become available. Wireless USB, which matches the same range but roughly the same 480Mbps peak speed of its wired equivalent. In new trends Gi-Fi wireless technology has been developed and can be replacement for technologies such as Bluetooth and ultra-wideband (UWB). The process of Gi-Fi would use a chip that transmits at an extremely high 60GHz frequency versus the 5GHz used for the fastest forms of Wi-Fi.

The sheer density of the signal would allow a chip to send as much as 5 gigabits per second. While the spectrum would limit the device to the same 33-foot range as Bluetooth or UWB, it could theoretically transfer an HD movie to a cell phone in seconds. Mixing and signal filtering used in Gi-Fi technology would keep the signal strong versus the longer-ranged but slower and more drop-prone Wi-Fi option of today. The chip in Gi-fi would likely cost is less.

Table.1. Comparison of GI-FI and Existing Technologies

| Characteristics | Bluetooth | Wi-Fi | Gi-Fi |
|---|---|---|---|
| Specification Authority | Bluetooth SIG | IEEE, WECA | NICTA |
| Development Start date | 1998 | 1990 | 2004 |
| Primary device | Mobile phones, PDAs, Consumer, Electronics office Industrial Automation Devices | Notebook, Computers, Desktop, Computer servers | Mobile phones, Home devices, PDAs, Consumer, Electronics office Industrial Automation Devices |
| Power consumption | 5mw | 10mw | < 2mw |
| Data transfer rate | 800Kbps | 11Mbps | 5 Gbps |
| Range | 10 meters | 100 meters | 10 meters |
| Frequency | 2.4GHz | 2.4GHz | 57-64GHz |

## 5. Benefits of GI-FI Technology

The most important benefits of the Gi-Fi technology are as follows:

5.1 Removing Cables

For many years cables ruled the world. Optical fibers played a dominant role for its higher bit rates and faster transmission. But the installation of cables caused a greater difficulty and thus led to wireless access. The standard's original limitations for data exchange rate and range and high cost of the infrastructures have not yet made it possible for Wi-Fi to become a good replace for the cables. Gi-Fi technology Removes need for cables to connect consumer electronics devices and all the devices can be connected in order to transmit the data wirelessly.

5.2 Cost of Chip is low

Gi-Fi's chip uses only a tiny one-millimeter-wide antenna and less than 2mili watts of power. Low-cost chip allows technology to be readily incorporated into multiple devices. The chip in Gi-fi would likely cost less to build . Then a small design would allow cell phones and other small devices to add the technology without significantly drive up the price. Gi-Fi is based on an open, international standard. Mass adoption of the standard, and the use of low-cost, mass-produced chipsets, will drive costs down dramatically, which is very less in compare to present technologies.

5.3 Privacy and Security

Encryption technology in Gi-Fi ensures privacy and security of content. About 70 per cent of firms have deployed their WLAN in a secure firewall zone but are still using the old WEP protocol, which does not protect the application layer effectively, so better encryption is urgently needed.

### 5.4 Flexibility

One of the problems with wire connections and cables is complexity for connecting, but in the Gigabit wireless technology simplicity is one of the features. Simple connection improves the consumer experience. The benefits related to the Gi-fi technology that can be achieved by the deployment and use of this technology.

## 6. Conclusion

Gi-Fi has given and it is conspicuous that more research should be done in the field of this new wireless technology and its applications .The Bluetooth which covers 9-10mts range and wi-fi followed 91mts .no doubt introduction of wi-fi wireless network has proved a revolutionary solution to bluetooth problem  the standard original limitations for data exchange rate and range, number of chances, high cost of infrastructure have not yet possible for wi-fi to become a power network, then towards this problem the better technology despite the advantages of rate  present technologies led to the introduction of new ,more up to date for data exchange that is GI-FI. The comparison is performed between Gi-Fi and existing wireless technologies in this paper shows that these features along with some other benefits that make it suitable to replace the existing wireless technologies. It removes cables that for many years ruled over the world and provides high speed data transfer rate. Gi-Fi technology has much number of applications and can be used in many places and devices such as smart phones, wireless pan networks, media access control and mm-Wave video-signals transmission systems.

**Jyoti Tewari-** B.Tech in CSE  in 2012, pursuing Mtech in CSE ,Research interest is in network and security. Two papers works is going on under the upcoming conferences. The book related work is going on.

**Swati Arya-** B.Tech in Computer Engineering in 2012, pursuing M.Tech in CSE. Research Interest is in wireless networking. Paper work is going through upcoming conferences. One Paper has been published in CACCS-2013.